\documentclass[aps,prl,floatfix,twocolumn,amsmath,amssymb]{revtex4-1}
\usepackage{amsthm,color,latexsym,graphicx,array,multirow,epstopdf}
\usepackage[hyperfootnotes=true]{hyperref}

\newcommand{\nn}{\nonumber}

\newcommand{\eq}[1]{Eq.~\eqref{eq:#1}}
\newcommand{\df}{\mathrm{d}}
\newcommand{\rd}{\mathrm{d}}

\newcommand{\al}{\alpha}

\newcommand{\de}{\delta}

\newcommand{\si}{\sigma}

\newcommand{\cG}{{\mathcal G}}
\newcommand{\cJ}{{\mathcal J}}

\newcommand{\Q}{ {\mathcal{Q}}}
\newcommand{\Dt}{ {\widetilde{D}}}
\newcommand{\Qk}{\Q_\kappa^i}

\newcommand{\jet}{\mathrm{jet}}

\begin{document}

\title{Jet Charge at the LHC}

\author{David Krohn}
\email{dkrohn@physics.harvard.edu}
\affiliation{Department of Physics, Harvard University, Cambridge MA, 02138 \vspace{-1ex}}
\author{Tongyan Lin}
\email{tongyan@uchicago.edu}
\affiliation{Kavli Institute for Cosmological Physics, Enrico Fermi Institute, University of Chicago, Chicago, IL 60637 \vspace{-1ex}}
\author{Matthew D. Schwartz}
\email{schwartz@physics.harvard.edu}
\affiliation{Department of Physics, Harvard University, Cambridge MA, 02138 \vspace{-1ex}}
\author{Wouter J. Waalewijn}
\email{wouterw@physics.ucsd.edu}
\affiliation{Department of Physics, University of California at San Diego, La Jolla CA, 92093 \vspace{-0.5ex}}

\date{\today}

\begin{abstract}
Knowing the charge of the parton initiating a light-quark jet could be
extremely useful both for testing aspects of the Standard Model and
for characterizing potential beyond-the-Standard-Model signals. We show that despite
the complications of hadronization and out-of-jet radiation such as pile-up, 
a weighted sum of the charges of a jet's constituents can
be used at the LHC to distinguish among jets with different
charges.  Potential applications  include
measuring electroweak quantum numbers of hadronically
decaying resonances or supersymmetric particles,
as well as Standard Model tests, such as jet
charge in dijet events or in hadronically-decaying $W$
bosons in $t\bar{t}$ events.  We develop a systematically improvable method to calculate
moments of these charge distributions by combining multi-hadron
fragmentation functions with perturbative jet functions and pertubative
evolution equations.  We show that the dependence on
energy and jet size for the average and width of the jet charge can be
calculated despite the large experimental uncertainty on fragmentation
functions. These calculations can provide a validation tool for data independent of
 Monte-Carlo fragmentation models.
\end{abstract}
\maketitle

The Large Hadron Collider (LHC) at CERN provides an opportunity to
explore properties of the Standard Model in unprecedented detail
and to search for physics beyond the Standard Model in previously unfathomable ways.
The exquisite detectors at {\sc atlas} and {\sc cms}
let us go beyond treating jets simply as 4-momenta to treating them as 
objects with substructure and quantum numbers.
A traditional example is whether a jet was
likely to have originated from a $b$-parton. At the LHC, one can additionally 
explore whether a jet has subjet constituents,
as from a boosted heavy object decay~\cite{Butterworth:2008tr,Kaplan:2008ie}, 
or whether it originated from a quark or gluon~\cite{Gallicchio:2011xq}. 
See Ref.~\cite{Altheimer:2012mn} for a recent review of jet substructure.
Here we consider the feasibility of measuring the electric charge of a jet.

The idea of correlating a jet-based observable to the charge of the
underlying hard parton has a long history. In an effort to determine
the extent to which jets from hadron collisions were similar to jets
from leptonic collisions, Field and Feynman~\cite{Field:1977fa} argued
that aggregate jet properties such as jet charge could be
measured and compared. The subsequent measurement at
Fermilab~\cite{Berge:1980dx} and CERN~\cite{Albanese:1984nv} in
charged-current deep-inelastic scattering experiments
showed clear up- and down-quark jet discrimination, 
confirming aspects of the parton
model. Another important historical application was the light-quark
forward-backward asymmetry in $e^+e^-$ collisions, a precision
electroweak observable~\cite{Decamp:1991se}.  Despite its historical
importance, there seem to have been no attempts yet at measuring
the charge of light-quark jets at the LHC.

Most experimental studies of jet charge measured variants of a
momentum-weighted jet charge. We define the $p_T$-weighted jet charge for
a jet of flavor $i$ as 
\begin{equation} \label{Omdef}
    \Qk = \frac{1}{(p_T^{\rm jet})^\kappa} \sum_{j \in \mathrm{jet}} Q_j (p_T^j)^\kappa
\end{equation}
where the sum is over all particles in the jet,
$Q_j$ is the integer charge of the color-neutral object observed, $p_T^j$ is the magnitude of its transverse
momentum with respect to the beam axis, $p_T^{\rm jet}$ is the total transverse momentum of the jet, and $\kappa$ is a free parameter. A common variant uses energy instead of $p_T$.  
Values of $\kappa$ between 0.2 and 1 have been used in experimental studies~\cite{Berge:1980dx,Decamp:1991se}.
 
\begin{figure}[b]
 \includegraphics[width=0.45\textwidth]{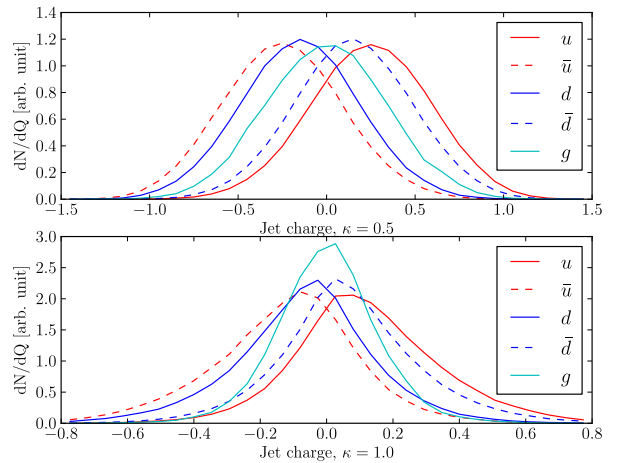}
\caption{Distributions of $\Qk$ for various parton 
flavors obtained from $pp\to W'\to \bar{q}q$ or $pp\to gg$ events with $p_T^{\rm jet} = 500$ GeV and $\kappa=0.5,1$. 
\vspace{-2ex}
 \label{fig:ex_dist}}
 \end{figure}

In hadron-hadron collisions at high energy, such as at the LHC, the particle multiplicities in 
the final state are significantly larger than at low energy and at $e^+e^-$ or lepton-hadron colliders.  
Thus, one would expect that measuring the
charge of a light-quark jet at the LHC should be 
difficult, with the primordial quark charge quickly getting washed out.  
However, this turns out not to be the case. For example,
Fig.~\ref{fig:ex_dist} shows distributions of $\Qk$ for $u,\bar{u}, d, \bar{d}$ and $g$
jets for two values of $\kappa$~\footnote{A note on simulation: Our Monte-Carlo events are
  generated in {\sc Madgraph 5}~\cite{Alwall:2011uj}, showering and hadronization are modeled using {\sc
    pythia 8}~\cite{Sjostrand:2007gs}, and jets are clustered with
  {\sc Fastjet}~\cite{Cacciari:2011ma} with
  anti-kT~\cite{Cacciari:2008gp} jets of $R=0.5$. Unless stated
  otherwise, we  assume a 14 TeV LHC.}. One can
clearly see that $\Qk$ will be useful for
identifying the charge of the primordial parton.
Moreover, as we will show, the energy and
and jet-size dependence of moments of jet-charge distributions can be
calculated in perturbative QCD.
 
 \begin{figure}[t]
 \includegraphics[width=0.45\textwidth]{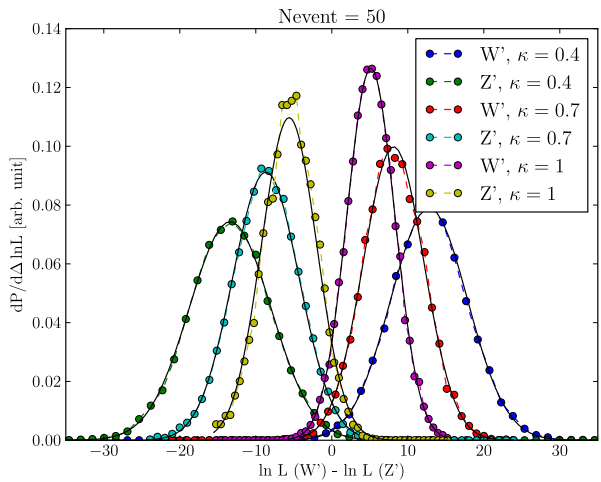}
\caption{Distinguishing $W'$ from $Z'$ with a log-likelihood
  discriminant, for different values of $\kappa$. Even with only 50
  events the samples are extremely well separated. \vspace{-2ex}
\label{fig:WprimeZprime}}
 \end{figure}

To get an impression of how much data is needed for $\Qk$ to be useful, we
consider measurements designed to distinguish charged from neutral vector 
resonances. To be concrete, we consider
scaled-up $W$ and $Z$ bosons at a mass of 1 TeV decaying into light
quark jets.  Simply 
cutting on the sum of the $\Qk$ of the hardest two jets in each event 
we can distinguish the two samples (assuming no background) with $95\%$ confidence
using around 30 events.  We find that the best discriminating power is 
achieved for $\kappa\sim0.3$.  A more sophisticated log-likelihood 
discriminant based on the two-dimensional jet charge distribution is shown in
Fig.~\ref{fig:WprimeZprime}, where $\sim 4\sigma$ separation of
the two samples is achievable with 50 events.

For another phenomenologically relevant application of jet charge
consider a simplified supersymmetric model with
squarks pair produced through $t$-channel gluino exchange and
 decaying as $\tilde{q}\rightarrow q+\chi_0^1$.  
 At $m_{\tilde{q}}=m_{\tilde{g}}=1.5~{\rm TeV}$ such a
model is still allowed~\cite{:2012rz}, although it will come under
scrutiny with the next round of 8 TeV data.  Due to the high concentration 
of up-type valence quarks at large $x$, the di-squark production process yields many events with
two hard up-type jets and missing energy, in contrast to the
background (dominated by $V$+jets) where the two hardest jets are
rarely both ups.  Adopting a set of cuts similar to those of Ref.~\cite{:2012rz}, 
we estimate if an excess is seen in 2 jets and missing energy channel, the 
increased concentration of up quarks could be measured above the $2\sigma$ level with
$25~{\rm fb}^{-1}$ of 8 TeV data, providing unique insights into the flavor structure of the new physics.
%
%

\begin{figure}[t]
 \includegraphics[width=0.4\textwidth]{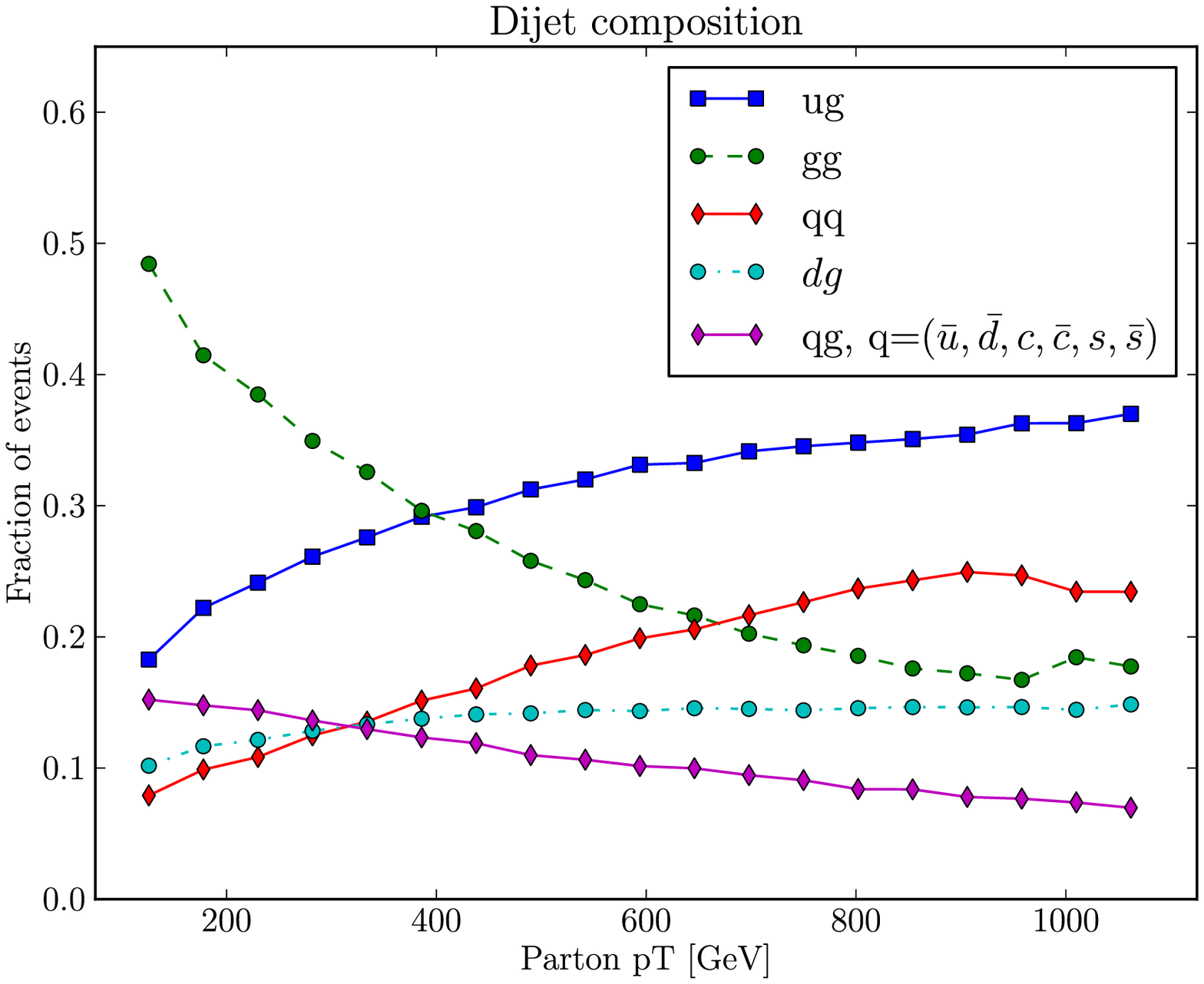}
\includegraphics[width=0.4\textwidth]{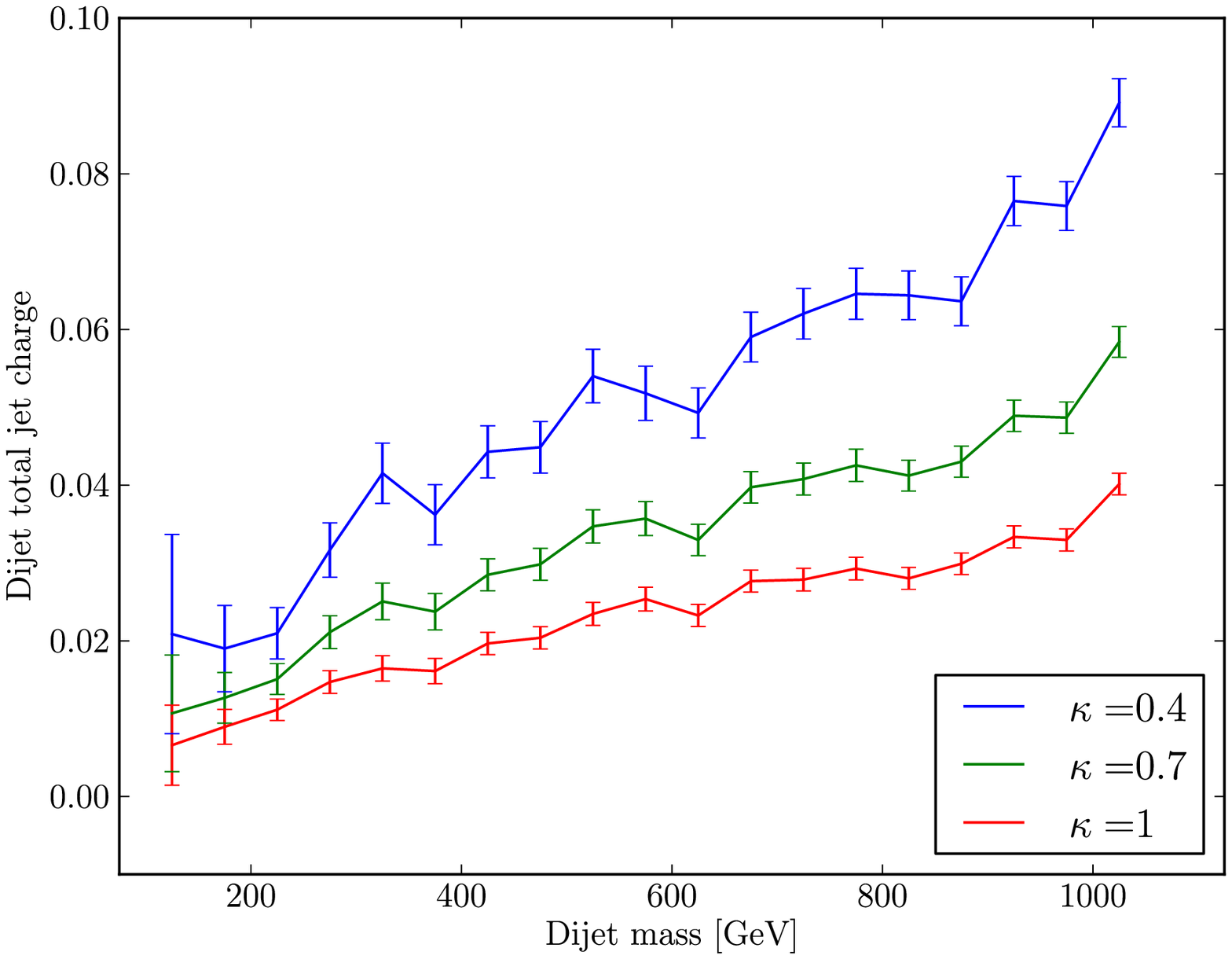}
\caption{Top: final state composition in dijet production.
Bottom: Sum of the two jet charges in dijet events, for various $\kappa$. The
  growth with dijet invariant mass reflects the larger fraction of
  valence quark PDFs at large $x$ and corresponding decrease in $gg$
  final states. \vspace{-3ex}
  \label{fig:dijetchargept}}
 \end{figure}

 \begin{figure}[t]
 \includegraphics[width=0.45\textwidth]{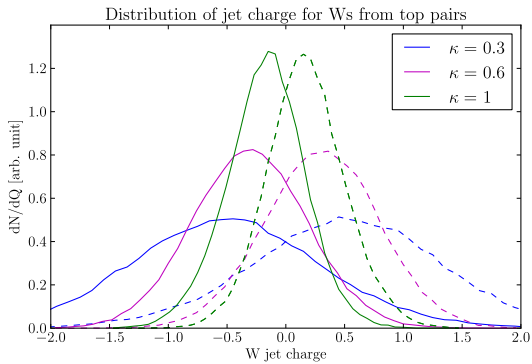}
\caption{Sum of jet charges of the
  two non $b$-jets in semi-leptonic $t\bar{t}$ events with a
  positively (solid) or negatively (dashed) charged lepton. \vspace{-2ex}
\label{fig:topWs}}
 \end{figure}

To trust a measurement of jet charge, it is important 
to test it on samples of known composition. While
proton collisions do not generally provide clean samples of pure
up- or down-quark jets, there are still ways to validate the
method on data.  For example, dijet production has an enormous cross
section at the LHC and the fraction
of jets originating from different partons is directly determined by
the parton distribution functions (PDFs).
At larger energies the valence quark PDFs 
dominate over gluon or sea quark PDFs, producing
more charged final states, as can be seen in 
see Fig.~\ref{fig:dijetchargept}.  The mean total jet charge in dijet 
events is also shown 
for various values of $\kappa$.  Verifying the
trend in this plot on LHC data would help validate jet charge.

Another sample of interest for validating jet charge
is hadronically decaying $W$ bosons 
coming from top decays.  In a semi-leptonic $t\bar{t}$
sample, the leptonically decaying $W$ can be used to determine the
two charges of the jets from the hadronically decaying $W$. 
The distributions of these charges can then be 
compared to expectations, an example comparison is shown in
Fig.~\ref{fig:topWs}. Validating this simulation on data would
establish weighted jet charge as a trustworthy tool, which could then
be used for new physics applications.  Perhaps it could even be
employed within the context of $W$ decays to help with top-tagging or
$W$ polarization measurements.

\begin{figure}[t]
 \includegraphics[width=0.45\textwidth]{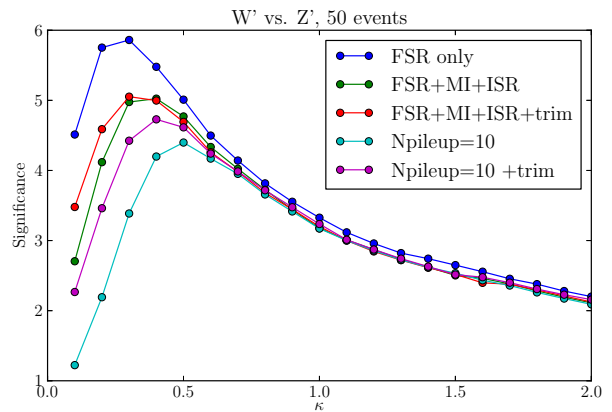}
\caption{Comparison of $W'$ vs. $Z'$ discrimination subject to
  contamination from  initial state radiation
  (ISR), multiple interactions (MI), and pile-up
  events. We also show the result with and without jet trimming
  ($R_{\rm sub}=0.2$, $f_{\rm cut} = 0.02$). \vspace{-2ex}
\label{fig:contamination}}
 \end{figure}

Next, we consider the effects of pile-up and
contamination on jet charge.  One might worry that at high luminosity
jet charge would be diluted by pile-up events, as up to ${\cal
  O}(100)$ proton-proton collisions can take place in the same bunch
crossing.  However, the products of these interactions tend to be
soft, and are thus assigned little weight as long as $\kappa$ 
is not too small. Further, charged particles can be traced to their collision vertex
allowing most contamination to be removed.
Finally, jet grooming techniques like trimming~\cite{Krohn:2009th} can be applied
to further reduce contamination.  We present a comparison of effects
of contamination and techniques to mitigate it in
Fig.~\ref{fig:contamination}.

Having demonstrated the practicality of jet charge for new physics
searches and proposed ways to validate it on standard model data, 
we now turn to the feasibility of systematically
improvable jet charge calculations. While Monte-Carlo programs like 
{\sc pythia} often provide an excellent approximation to full quantum 
chromodynamics, they are only valid to leading-order in perturbation
theory including the resummation of leading Sudakov double 
logarithms~\footnote{
Note that in addition Monte-Carlo programs differ in their formulation
of the parton shower and in their treatment of hadronization. 
Comparing {\sc pythia} to {\sc herwig++}~\cite{Gieseke:2011na}, we 
find agreement for the mean jet charge at the ${\cal O}(10\%)$ level.}.

A precise calculation of jet charge is challenging because it is not an
infrared-safe quantity. Jet charge is sensitive to hadronization 
and cannot be calculated without knowledge of the fragmentation functions
$D_j^h(x,\mu)$. These functions give the average probability that a hadron 
$h$ will be produced by a parton $j$ with the hadron having a fraction $z$ 
of the parton's energy. Fragmentation functions, like parton distribution 
functions, are non-perturbative objects with perturbative evolution equations which simplify in moment space. The Mellin
 moments are defined by
\begin{equation} \label{mellin}
 \Dt_q^h (\nu,\mu)= \int_0^1 \rd x\, x^\nu D_q^h(x,\mu)
\,,\end{equation}
which evolve through local renormalization group equations, just like 
the moments of parton distribution functions.

We first consider the average value of the jet charge
\begin{equation}
 \!\langle \Qk \rangle = \frac{1}{\sigma_\jet}\int\! \rd \sigma\, \Qk 
  = \int\! \rd z\, z^\kappa \sum_h Q_h\, \frac{1}{\si_\text{jet}} \frac{\df \si_{h \in \text{jet}}}{\df z}
\,,\end{equation}
where $z=E_h/E_\jet$ is the fraction of the jet's energy the hadron carries. 
For narrow jets $z\sim p_T^h/p_T^\jet$. 

To connect to the fragmentation functions, we first observe
that for $\kappa > 0$ the the charge is dominated by collinear 
and not soft radiation. Thus the contributions of the hard and soft sectors
of phase space, while contributing to the formation of the jet, should have 
a suppressed effect on $\Qk$. We can therefore use the fragmenting jet 
functions introduced in Refs.~\cite{Procura:2009vm,Procura:2011aq} to write
\begin{equation}
\frac{1}{\sigma_\text{jet}}
 \frac{\df \si_{h \in \text{jet}}}{\df z}
= \frac{1}{16\pi^3} \sum_j \int_z^1 \frac{\rd x}{x}  
\frac{\cJ_{i j}(E,R,\frac{z}{x},\mu)}
{\cJ_i(E,R,\mu)} D_j^h(x,\mu)
\,.\end{equation}
Here $\cJ_i(E,R,\mu)$ is a jet function and $\cJ_{ij}(E,R,x,\mu)$ a set of calculable
coefficients which depend on the jet definition and flavor $i$ of the hard 
parton originating the jet. The hard and soft contributions conveniently 
canceled in this ratio. Therefore
\begin{equation} \label{aveq}
 \langle \Q_\kappa^q \rangle = \frac{1}{16\pi^3} \frac{\widetilde{\cJ}_{q q}(E,R,\kappa,\mu)}
{\cJ_q(E,R,\mu)} \sum_h Q_h \Dt_q^h(\kappa,\mu)
\,,\end{equation}
with $\widetilde{\cJ}_{i j}$ related to ${\cJ}_{i j}$ by a Mellin-transform as 
in Eq.~\eqref{mellin}. By charge conjugation $\sum_h Q_h \Dt_g^h(\kappa,\mu)=0$, 
so in particular $\langle \Q_\kappa^g \rangle = 0$. We have checked that the 
$\mu$-dependence of $\cJ_{ij}/\cJ_i$ exactly compensates for the 
$\mu$-dependance of the fragmentation functions at order $\al_s$.

We have written both $\cJ_i(E,R,\mu)$ and $\cJ_{ij}(E,R,x,\mu)$ as if they 
depend on the energy $E$ and size $R$ of the jet, however, these functions 
only give a valid description to leading power of a single scale corresponding 
to the transverse size of the jet. Here we use the $e^+e^-$ version of anti-$k_T$
jets of size $R$, for which the natural scale is  $\mu_j = 2E \tan (R/2)$~\cite{Ellis:2010rwa}.
We can therefore calculate the average jet charge by evaluating the Mellin-moments
of fragmentation functions at the scale $\mu_j$ and multiplying by the jet functions.

\begin{figure}[t]
 \includegraphics[width=0.4\textwidth]{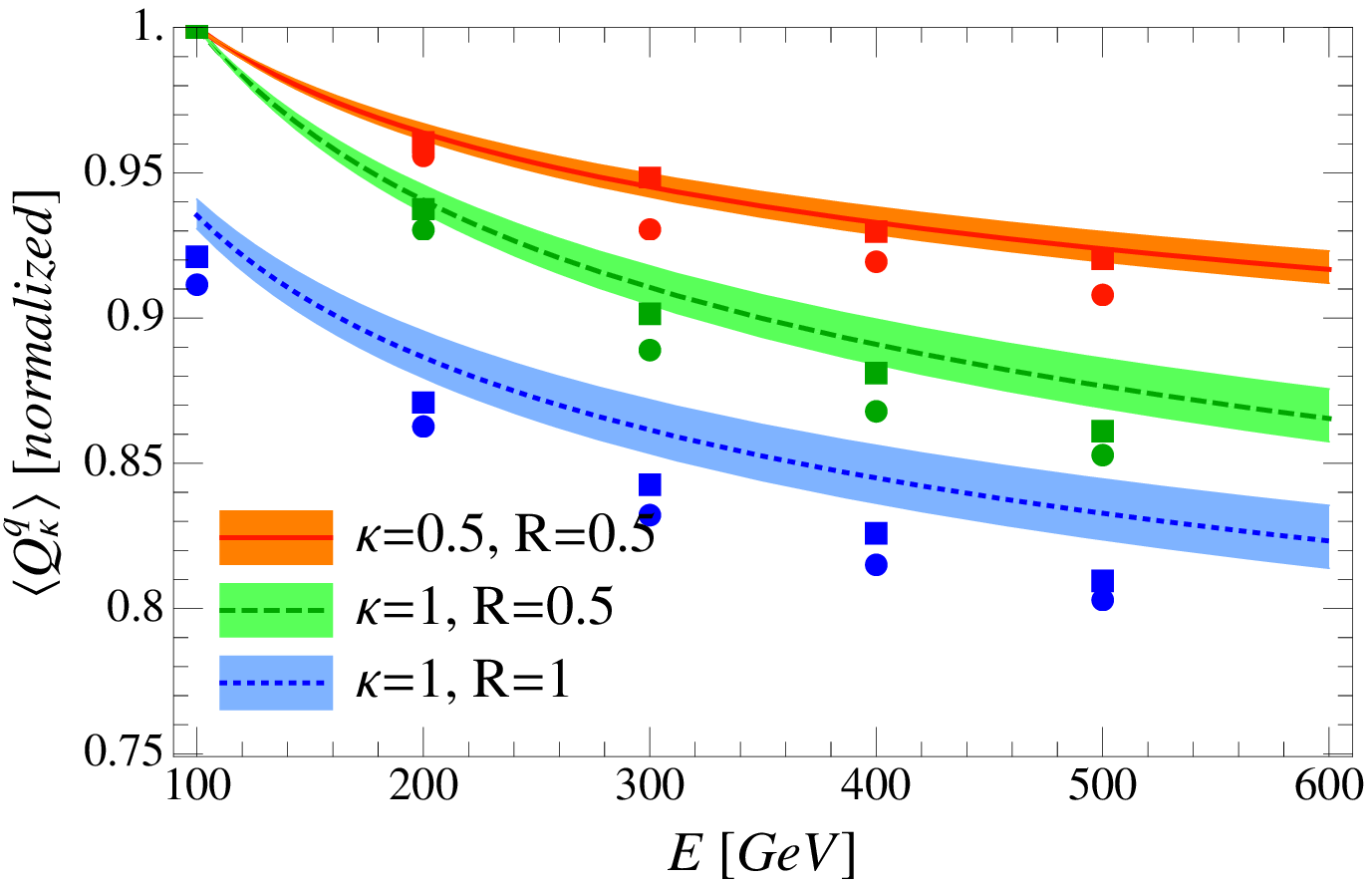}
 \includegraphics[width=0.4\textwidth]{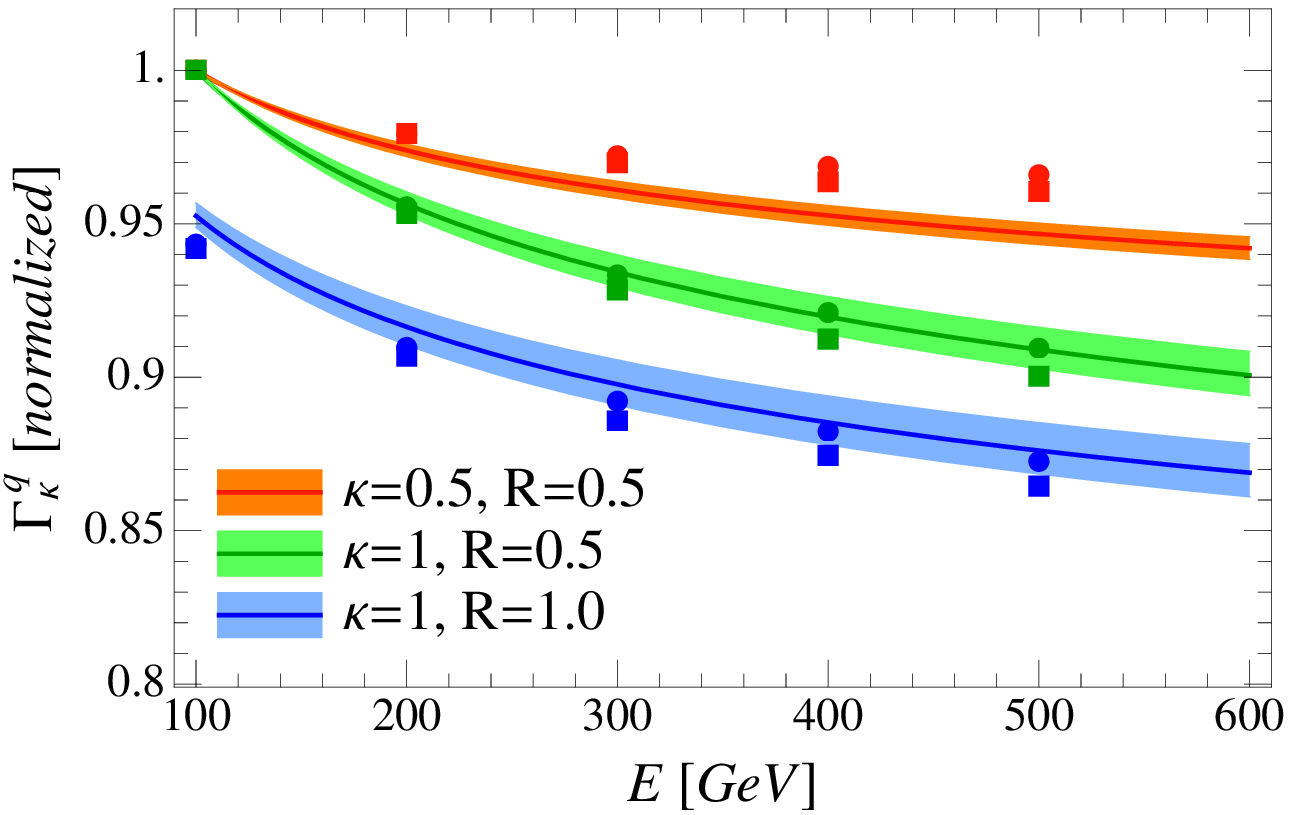}
\caption{Comparison of theory prediction  (bands) for the average (top) and width (bottom) of the jet charge distribution to {\sc pythia} (squares and circles for $d$ and $u$ quarks)
for $e^+e^-$ collisions. Bands show uncertainty from varying the factorization scale by a factor of $2$. 
Normalizing to 1 at $E=100$ GeV and $R=0.5$ removes the dependence on the nonperturbative input and quark flavor.
%
\label{fig:avwid}}
\end{figure}

Since only one linear combination of fragmentation functions appears
in Eq.\eqref{aveq}, the theoretical prediction is not significantly limited 
by the large uncertainty on $D_j^h(\kappa,\mu)$. One can simply measure 
$D_j^h(\kappa,\mu)$ by observing the average jet charge for each flavor at 
one value for $\mu$ and then using the theoretical calculation to predict it at 
other values. In the absence of data, we simulate such a comparison using 
{\sc pythia}. The result is shown in Figure~\ref{fig:avwid} for various values 
of $\kappa$ and $R$, and normalized at a reference point. Already we can 
see a clear agreement between the theory and {\sc pythia}.

%

To calculate other properties of the jet charge distribution requires
correlations among hadrons. For example, we can consider the width of the 
jet charge, $(\Gamma_\kappa^i)^2 = \langle \Qk \rangle^2 -\langle (\Qk)^2 \rangle $. 
This depends on the moment
\begin{align}\label{width}
   \big\langle (\Qk)^2 \big\rangle 
&=\sum_n \sum_{h_1,\ldots,h_n}  \int\! \df z_1\cdots \df z_n\, (Q_1 z_1^\kappa+\cdots +Q_n z_n^\kappa)^2
\nn \\
&\qquad \times \frac{1}{\si_\text{jet}}\, \frac{\df^n \si_{h_1 \cdots h_n \in \text{jet}}}{\df z_1\cdots \df z_n} 
\,,\end{align}
where the sum runs over all hadronic final states. After integrating over most 
of the $z_i$ and including a factor of $\frac{1}{2}$ for identical hadrons, 
this simplifies to 
\begin{align}
   \big\langle (\Qk)^2 \big\rangle 
&= \int\! \df z\, z^{2\kappa} \sum_h Q_h^2 \frac{1}{\si_\text{jet}} \frac{\df \si_{h \in \text{jet}}}{\df z} \\ 
&\quad + \int\! \df z_1\, \df z_2\, z_1^\kappa z_2^\kappa \sum_{h_1,h_2} Q_{h_1} Q_{h_2}  \frac{1}{\si_\text{jet}} \frac{\df \si_{h_1 h_2 \in \text{jet}}}{\df z_1\, \df z_2} \nn
\,.\end{align}  
The first term on the right hand side can be expressed in terms of products of fragmentation functions and jet
functions as for $\left\langle \Qk \right\rangle$. The second term can be expressed in terms of something 
we call a dihadron fragmenting jet function, $\cG_i^{h_1h_2}$. Its matching onto (dihadron) fragmentation functions is given by
\begin{align} \label{eq:Ghh}
& \cG_i^{h_1 h_2}(E,R,z_1,z_2,\mu) 
 \\
&= \sum_j \int\! \frac{\df u}{u^2} \,\cJ_{ij}(E,R,u,\mu) D_j^{h_1 h_2}\Big(\frac{z_1}{u},\frac{z_2}{u},\mu\Big) \nn \\
 & + \sum_{j,k} \int\! \frac{\df u}{u} \frac{\df v}{v} \,\cJ_{ijk}(E,R,u,v,\mu) D_j^{h_1}\Big(\frac{z_1}{u},\mu\Big) D_k^{h_2}\Big(\frac{z_2}{v},\mu\Big)
\nn\,,\end{align}
The second term is due to a perturbative parton splitting before hadronization 
and only starts at 1-loop order,
\begin{align}
 \cJ_{ijk}^{(1)}(E,R,u,v,\mu) =  \cJ_{ij}^{(1)}(E,R,u,\mu) \de(1\!-\!u\!-\!v) \de_{k,a(ij)}
\,,\end{align}
where $\de_{k,a(ij)}$ indicates that the flavor $k$ is completely fixed by $ij$. 
E.g.~$a(qq)=g$, $a(gq) =\bar q$. We then find
\begin{align} \label{Qsig}
 \big\langle  (\Q_\kappa^q)^2 \big\rangle 
& = \frac{1}{16\pi^3} \sum_j \frac{\widetilde \cJ_{qj}(E,R,2\kappa,\mu)}{J_q(E,R,\mu)}
 \Big[\sum_h Q_h^2 \tilde D_j^h(2\kappa,\mu)
\nn \\ & \quad 
 + \!\!\sum_{h_1,h_2}\! Q_{h_1} Q_{h_2} \tilde D_j^{h_1 h_2}(\kappa,\kappa,\mu) \Big]
\,.\end{align}  
(For a gluon jet, which we do not consider here, there is a contribution from the 
last line of \eq{Ghh} corresponding to a perturbative $g \to q\bar{q}$ splitting.)
We have checked that this equation is renormalizat-group invariant at order $\al_s$.

Unfortunately, the dihadron fragmentation functions are even more poorly known
than the regular fragmentation functions. However, we can use the same trick as 
for the average jet charge to calculate the $E$ and $R$ dependence of the width, 
given measurements at some reference scale.  As with the average jet charge,
we can now calculate the width by fitting one parameter for each flavor, corresponding 
to the term in brackets in Eq.~\eqref{Qsig}, and predicting the $E$ and $R$ dependence. 
Results compared to {\sc pythia} for the width are shown in Fig.~\ref{fig:avwid} and
show good agreement.
The gluon mixing contribution is not included in these figures since it requires additional matching; a discussion of the effect of gluon mixing can be found in Ref.~\cite{Waalewijn:2012sv}.

To go beyond the average and the width, for example to the 3rd or higher moments, 
multi-hadron fragmentation functions would be needed. From a practical point of view, 
such functions are nearly impossible to measure with any precision. However, we have 
found that the discriminating power of jet charge is nearly as strong 
using Gaussians based on the average and width as it is with
the full differential jet charge distribution. It follows that accurate calculations of the 
phenomenologically relevant part of jet charge distributions are achievable with the 
formalism we have introduced in this paper. The full fragmenting jet functions, both 
for the single hadron and dihadron case, and the evolution kernels, are now known
at 1-loop order. To see whether higher precision is required, and to explore
the importance of power corrections, requires some LHC data to compare with. 
The calculations and issues discussed here are expanded on in Ref.~\cite{Waalewijn:2012sv}.

As we have shown, the weighted jet charge, and its moments, are measurable 
and testable already at the LHC. With potential to uniquely determine quantum 
number of certain new physics particles, should they show up, it is important to 
verify jet charge on standard model processes. Thus jet charge holds promise 
as a measurable, calculable and useful observable.


We thank G. Kane, E. Kuflik, S. Rappoccio, G. Stavenga, and M. Strassler for helpful discussions.
 DK is supported by the Simons foundation and by an LHC-TI travel grant.  TL is partially supported by NASA Theory Program grant NNX10AD85G and by the National Science Foundation under Grant No. PHYS-1066293 and the hospitality of the Aspen Center for Physics. MDS is supported by DOE grant DE-SC003916.  WW is supported by DOE grant DE-FG02-90ER40546. 


\bibliography{jetcharge}
\end{document}